\documentclass[twocolumn,showpacs,aps,pre]{revtex4}
\usepackage{amssymb,amsmath,epsfig,graphicx}
\usepackage{mathrsfs}

\newcommand{\Sinfo}{\Delta S_{\rm info}}
\newcommand{\past}[1]{\mathbf{C}^{#1}}

\begin{document}
\title{Thermodynamics of feedback controlled systems}
\author{F. J. Cao$^{1,2,}$}
\email{francao@fis.ucm.es}
\author{M. Feito$^{1,}$}
\email{feito@fis.ucm.es}
\affiliation{$^1$ Departamento de
F\'{i}sica At\'omica, Molecular y Nuclear, Universidad Complutense
de Madrid, Avenida Complutense s/n, 28040 Madrid, Spain}
\affiliation{$^2$ LERMA,
Observatoire de Paris, Laboratoire Associ\'e au CNRS UMR 811 2,
61, Avenue de l'Observatoire, 75014 Paris, France.}

\begin{abstract}
We compute the entropy reduction in feedback controlled systems due to
the repeated operation of the controller. This was the lacking ingredient
to establish the thermodynamics of these systems, and in
particular of Maxwell's demons. We illustrate some of the consequences of
our general results by deriving the maximum work that can be extracted from
isothermal feedback controlled systems. As a case example, we finally study
a simple system that performs an isothermal
information-fueled particle pumping.
\end{abstract}
\pacs{89.70.Cf, 05.20.-y}
\maketitle

\section{Introduction}

Controllers are ubiquitous in science and
technology with a number of purposes such as stabilizing  unstable dynamics or
increasing the performance~\cite{bec05}. Furthermore, many real systems in
nature can be modeled as a system plus a controller.
A controller is an external agent whose action is to modify the evolution of
the system with a purpose.
\emph{Feedback or closed-loop} controllers use information about the state of
the system.
The feedback is the process performed by the controller of measuring the
system, deciding on the action given the measurement output, and acting
on the system.
On the contrary, an open-loop controller
operates on the system blindly, i.e., without information of its
state. Although it is intuitively clear that the
information about the state of the system can be used to improve
the performance, there are still open questions on the
connections between feedback control theory and information theory
(see Ref.~\cite{bec05}). In particular, the understanding of the
thermodynamics of feedback control is still incomplete. Much of
the progress in the solution of this problem has come from the study of
Maxwell's demon~\cite{lef03}. This is a being that gathers
information about a system and is able to decrease the entropy of the
system without performing work on it. The seminal work of Szilard~\cite{szi29}
contains the basic ingredients of the trade off 
between information theory and thermodynamics, which is precisely
stated in Landauer's principle: The erasure of $1$ bit of
information at temperature $T$ implies an energetic cost of at
least $k_BT\ln 2$~\cite{lan61}. Bennett~\cite{ben82} pointed out
that Landauer's principle is the key to preserving the second law of
thermodynamics in feedback systems, as the controller must erase
its memory after each cycle to allow the whole system to truly
operate cyclically. How to achieve the shorter description for the
memory record of the controller in order to minimize the energetic
erasure cost was established by Zurek~\cite{zur89} by using an
algorithmic complexity approach. On the other hand, Lloyd used
in~\cite{llo89} a different point of view ---that of the feedback
controlled system. From this approach the effect of the
interaction of the controller with the system is to reduce the
entropy of the system, due to the additional determination of the
macrostate of the system through the information obtained from it.
More recently, Touchette and Lloyd~\cite{tou00} have computed the
maximum additional reduction in entropy attainable in one control
action when a feedback control is used instead of an open-loop
control.
\par

In this paper we also consider the point of view of the feedback
controlled system. The thermodynamics of the interactions of the
system with the controller and the environment are well studied
for the heat and work exchanges. However, a complete understanding
of the entropy reduction in the system due to its interaction with
the feedback controller is still lacking. We solve here this problem and
show how to compute this entropy reduction after one or several control
steps. This result allows
us to establish the thermodynamics of feedback controlled systems without
assuming Landauer's principle. Several
concepts and quantities defined in information theory~\cite{cov91}
emerge naturally as one computes this entropy reduction. For the
definition of the entropy we will use $k_B=1$ and natural logarithms. This
implies that the information quantities that naturally appear will be in
nats ($\ln 2\mbox{ nats}=1\mbox{ bit}$).
\par

In the next section we compute the entropy reduction in a general
feedback controlled system due to the repeated operation of the
controller. The result allows us to establish the thermodynamics
of feedback controlled systems. In Sec.~\ref{sec:isothermal}, we
illustrate some of the consequences of our general result by
deriving the maximum work that can be extracted from isothermal
feedback controlled systems. In Sec.~\ref{sec:markovian}, we show
the applicability and usability of the results in a simple
dynamical system, a Markovian particle pump that is able to
extract useful work from the entropy reduction due to the
information used by an external feedback controller. Finally, we
summarize the results of the paper in Sec.~\ref{sec:conclussions}.
\par

\section{Entropy reduction in feedback controlled systems}

Let us call $X_k:=X(t_k)$ the macrostate of a general dynamical
system at the $k$th control step of the controller (at time
$t_k$). In a feedback controlled system the control step involves
several operations by the controller: measuring the system,
deciding the control action to take given the measurement output,
and acting on the system following the selected control action.
Therefore, the control action is the modification of the evolution
of the system made by the external agent that we shall call the
controller. The controller can perform several control actions on
the system. By $ C_1 = c $ we denote that, at the first control
step, the controller has chosen to perform the action labeled by
$ c $. (It is \emph{not} a specification of the state of the
controller.) As the control actions are decided at their respective
control steps, $C_k$ represents only the decision taken at
the $k$th control step.

Initially the entropy of the system is $S_0$, which is fixed by
the probabilities $p_{X_0}(x)$ of each possible microstate $x$ at
time $t=0$. Subsequently, the system evolves with an entropy
change from $S_0$ to $S^b_1$, which is the entropy just
\emph{before} the first control step. It is given by
the statistical entropy
\begin{equation}
  S^b_1=-\sum_{x\in\mathscr{X} }p_{X_1}(x) \ln p_{X_1}(x) =: H(X_1),
\end{equation}
with $\mathscr{X}$ as the set of possible microstates of the system.
At time $t_1$ the controller measures the state of the system. The
result of this measurement determines, at least partially, the
action the controller will take. The additional information on the
system provided by the measure further determines the system
macrostate~\cite{llo89}, i.e., it defines a submacrostate that
contains only microstates compatible with the measured value.
However, from the point of view of the system, each set of
measurement outputs that leads to the same control action can be
considered as defining a single submacrostate of the system,
because the controller in its action on the system ignores the
differences inside these sets. Thus, if the measurement implies a
control action $C_1=c$, the entropy of the system decreases to
\begin{equation}
H(X_1|C_1=c):=-\sum_{x\in\mathscr{X}} p_{X_1|C_1}(x|c) \ln
p_{X_1|C_1}(x|c).
\end{equation}
Therefore, the average entropy \emph{after} the first control step can be
obtained by averaging over the set $\mathscr{C}$ of all possible control
actions,
\begin{equation}
  S^a_1=\sum_{c\in\mathscr{C} }p_{C_1}(c)H(X_1|C_1=c)=:H(X_1|C_1).
\end{equation}
Hence the average variation in the entropy at the first step is
\begin{equation} \label{SinfoOne}
    \Delta S_1 = S^a_1-S^b_1=H(X_1|C_1)-H(X_1)=: -I(X_1;C_1),
\end{equation}
i.e., it is the (minus) mutual
information~\cite{cov91} between $X_1$ and $C_1$.
\par

Let us describe one more step. Each of the previous
$|\mathscr{C}|$ submacrostates of the system with entropy
$H(X_1|C_1=c)$ evolves to give an entropy $H(X_2|C_1=c)$ just
before the second control step. Following the second control step,
each one of these submacrostates of the system give
$|\mathscr{C}|$ more submacrostates. The entropy of the system
given that $C_1=c$ and $C_2=c^\prime$ is $H(X_2|C_2=c^\prime,C_1=c)$.
Therefore, the average entropy of the system after the second step
is
\begin{equation}
\begin{split}
  S^a_2 & =
\sum_{c,c^\prime\in\mathscr{C}} p_{C_2C_1}(c^\prime,c)
H(X_2|C_2=c^\prime,C_1=c)\\
& = H(X_2|C_2,C_1),
\end{split}
\end{equation}
and the average variation in the entropy at this
second control step is $\Delta S_2
=S^a_2-S^b_2=H(X_2|C_2,C_1)-H(X_2|C_1)=-I(X_2;C_2|C_1)$.
This conditioning of the mutual information shows that the entropy of the
system is only reduced by the new information.
\par

Analogously we get for the average entropy reduction in the $k$th step
$\Delta S_k=-I(X_k;C_k|\past{k-1})$, where $\past{k-1}$ stands for
${C_{k-1},C_{k-2},\dots,C_1}$. Using the properties of mutual
information~\cite{cov91}, this average entropy reduction can be
written as
\begin{equation}\label{sk2}
  \begin{split}
    \Delta S_k & = -I(X_k;C_k|\past{k-1})=-I(C_k;X_k|\past{k-1})\\
    \;& = - H(C_k|\past{k-1})+H(C_k|\past{k-1},X_k).
  \end{split}
\end{equation}
\par

Finally, we find that the \emph{total average entropy reduction due to the
  information used} in $M$ control steps is $\Sinfo=\sum_{k=1}^{M}\Delta S_k$,
  i.e.,
\begin{equation}\label{Sinfo}
  \Sinfo=-\sum_{k=1}^{M}I(C_k;X_k|\past{k-1}).
\end{equation}
This general result indicates that this entropy reduction can be computed in
terms of the joint probabilities for the state of the system and the control
actions history.
Using Eq.~\eqref{sk2} and the chain rule for $H$ (see
Ref.~\cite{cov91}), we rewrite the last equation as
\begin{equation}\label{Sinfo2}
  \Sinfo=-H(\past{M})+\sum_{k=1}^{M} H(C_k|\past{k-1},X_k).
\end{equation}
Equation~\eqref{Sinfo}, or equivalently Eq.~\eqref{Sinfo2}, is a central
result of this paper. As a consistency check, note that for
open-loop controlled systems the controller acts independently of
the state of the system and it gets no information of it.
Thus, $H(C_k|\past{k-1},X_k)=H(C_k|\past{k-1})$, which gives
$\Sinfo=0$ after applying the chain rule in Eq.~\eqref{Sinfo2}, as
expected. Note also that the mutual information in Eq.~\eqref{Sinfo} between
the system and the control actions is conditioned by the past control
actions. This reflects that the correlations between measurements limit the
attainable entropy reduction. Therefore, the entropy reduction in $M$
consecutive measurements is equal or lower than in $M$ independent
measurements.
\par

\subsection{Deterministic feedback controllers}

A relevant class of closed-loop controllers is
\emph{deterministic feedback controllers}. For them the
control action is determined without uncertainty by the state of
the system and the control actions history. Therefore
\begin{equation}\label{error-free-cond}
  H(C_k|\past{k-1},X_k)=0,
\end{equation}
and the entropy reduction in Eq.~\eqref{Sinfo2} simplifies to
$\Sinfo=-H(\past{M})$, which can be computed by just using the
joint probability $p_{C_1,\dots,C_M}(c_1,\dots,c_M)$.
Consequently, the average entropy reduction after a large number
of control actions is given by the entropy rate $\bar
H(\mathscr{C})$ of the stochastic process describing the control
actions:
\begin{equation} \label{SinfoRate}
 \lim_{M\to \infty} \frac{\Sinfo}{M}
 = \lim_{M\to \infty}\frac{-H(\past{M})}{M}=:-\bar H(\mathscr{C}).
\end{equation}
For a system and control dynamics without explicit dependencies in
time, this average entropy reduction coincides with the asymptotic
entropy reduction in one step~\cite{cov91}, that is, $\lim_{M\to
\infty} \Sinfo/M=\lim_{M\to \infty} \Delta S_M$.
\par

\subsection{Non-deterministic feedback controllers}

Feedback controllers satisfying Eq.~\eqref{error-free-cond} are
error free. On the other hand, controllers affected by some
\emph{source of error} are common in real systems. In this case the
decorrelation between the control actions and the state of the
system reduces the attainable entropy reduction; see
Eq.~\eqref{Sinfo2}. For instance, consider a feedback controller
with two possible actions, say ``on'' and ``off'', for which the
system state and the previous control actions history determine
which one of the actions is taken with probability $1-\epsilon$.
For this system, $H(C_k|\past{k-1},X_k)=H_b(\epsilon)$, with
$H_b(\epsilon)$ as the binary entropy function
$H_b(\epsilon):=-\epsilon \ln
\epsilon-(1-\epsilon)\ln(1-\epsilon)$, and Eq.~\eqref{Sinfo2} gives
\begin{equation}\label{error}
  \lim_{M\to \infty} \frac{\Sinfo}{M}=-\bar H(\mathscr{C})+H_b(\epsilon).
\end{equation}
This shows that errors in the control operation limit the attainable entropy
reduction.
\par

\subsection{Discussion}

The new relation~\eqref{Sinfo} sets the entropy reduction in the
controlled system  due to the information used by the external
agent that operates on it. The reformulation of this relation as
Eq.~\eqref{Sinfo2} allows us to understand the average entropy
reduction per control step as two competing contributions: a
negative term accounting for the entropy rate of the control
actions, and a positive term accounting for the decorrelation
between the controller actions and the state of the system. This
decorrelation can arise, for instance, from errors in the
operation of the controller [see Eq.~\eqref{error}]. These new
relations, Eqs.~\eqref{Sinfo} and~\eqref{Sinfo2}, also show how
the past control action history must be taken into account to
avoid redundancy in the computation of the entropy reduction. They
are consistent with the Zurek's computational interpretation of
the controller as a memory record whose blocks occupied by past
measurements must be compressed before the erasure
process~\cite{zur89,cav90}. On the other hand, when only one
control step is considered, Eq.~\eqref{Sinfo} reduces to
Eq.~\eqref{SinfoOne}, which gives the well-known Landauer's
energetic cost due to information~\cite{lef03}, $k_BT I(X_1; C_1)$
(recovering units), also found for quantum systems \cite{sag08}.
\par

The statement of the entropy reduction in terms of the control
actions is an important point of this paper. It allows one to give a
reachable bound for the efficiency. (If the controller performs the same
action for two different measured values, the bound found for the efficiency
considering the entropy reduction in terms of the measure could be
nonreachable.) Note also that the overall reduction in the
entropy of the system due to feedback control is expressed in terms of physical
quantities and it can be computed without knowledge of
internal details of the controller. In addition, this approach also
allows one to compute the maximum entropy reduction attainable
with a nondeterministic feedback control, Eq.~\eqref{error},
giving a reachable bound.
\par

The entropy reduction in the system due to the information used by
the controller is a fundamental ingredient in the thermodynamics
of feedback controlled systems. It is the key to improving the
performance in these systems compared with their open-loop counterparts.
Once this entropy reduction is understood and we know how to
compute it [Eqs.~\eqref{Sinfo} or~\eqref{Sinfo2}], the
thermodynamics of feedback controlled systems is complete. In
particular, we show in the next section how to compute
thermodynamic relations for an \emph{isothermal feedback
controlled system}.
\par

\section{Application: Isothermal feedback controlled systems}\label{sec:isothermal}

We study in this section the implications of the previous results for the case
of an isothermal feedback controlled system.

A general isothermal feedback controlled
system is a system that is coupled to a feedback controller, to a
thermal bath of temperature $T$, and to another external system
on which it does work. When the system is operated
cyclically, the initial state is recovered after a cycle, and the
variations in internal energy and entropy of the system in the
cycle are zero. During such a cycle the system releases a quantity
of heat $Q$ to the thermal bath and does work $W$ on the
external system. The transfer of the internal energy of the
controller $\Delta U_{\rm cont}$ to the system is given by the
first law of thermodynamics,
\begin{equation} \label{isofirst}
\Delta U_{\rm cont} + Q + W = 0.
\end{equation}
On the other hand, the second law of thermodynamics gives
\begin{equation} \label{isosecond}
T \Delta S_{\rm cont} + Q \geq 0,
\end{equation}
with $\Delta S_{\rm cont}$ as the entropy increase in the controller.
Combining both relations we get the inequality
\begin{equation} \label{isocomb}
W \leq - \Delta U_{\rm cont} + T \Delta S_{\rm cont} = - \Delta F_{\rm
cont},
\end{equation}
where $\Delta F_{\rm cont}$ is the variation in the Helmholtz free energy of
the controller in the cycle. From this relation it is natural to define the
efficiency of a feedback controlled system as
\begin{equation}\label{eta_def}
\eta = \frac{W}{-\Delta F_{\rm cont}}.
\end{equation}
In addition, if the controller only interacts with the system and
without heat transfer, we have $\Delta S_{\rm cont} \geq -
\Sinfo$, i.e., the increase in entropy of the controller should be
greater than or equal to the reduction in the entropy of the system
due to the actions of the controller. This implies that the
maximum efficiency that can be attained with an isothermal
feedback controlled system is
\begin{equation}\label{eta}
\eta = \frac{W}{- \Delta U_{\rm cont} - T \Sinfo},
\end{equation}
where $W$ is the work extracted from the system, $-\Delta U_{\rm cont} $ is
the work done by the controller on the system, and $\Sinfo$ is the entropy
reduction in the system due to the information-dependent operation of the
controller, which can be computed with Eq.~\eqref{Sinfo}.
\par

\section{Example: Markovian particle pump} \label{sec:markovian}

\begin{figure}
    \includegraphics [scale=0.6] {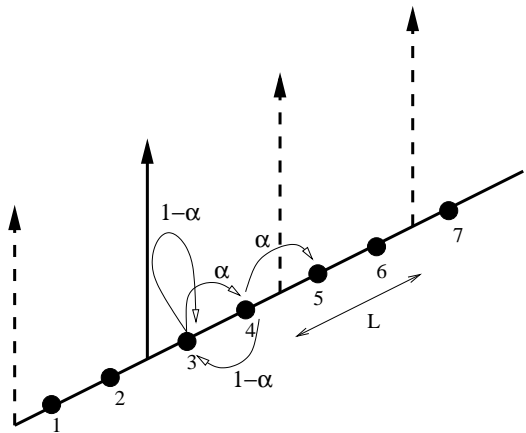}
    \caption{
Illustration of the Markovian particle pump with $n=2$ lattice sites between
barriers. This is a simple feedback controlled system that extracts useful
work from the entropy reduction due to the information about the system used
by the external feedback controller.
}
    \label{fig:figure}
\end{figure}

We shall illustrate how to apply our results in a simple dynamical
system, a \emph{Markovian particle pump}, which is able to
extract useful work from the entropy reduction due to the
information about the system used by an external feedback controller.
Consider a particle in a one-dimensional lattice that is in contact with a
thermal bath at temperature $T$. An external controller can
activate reflecting barriers separated by a distance $L$  with $n$ lattice
sites between two consecutive barriers; see Fig.~\ref{fig:figure}.
For the discussion of this example we will consider units of
$k_BT=1$ and $L=1$. In the absence of external forces, the particle
jumps to the left or to the right site with the same probability,
$1/2$, at each time step. Now let us have a force $f$ pointing in
the negative direction. The probability of jumping to the right
decreases and becomes $\alpha:=1/(1+e^{f/n})$, as follows from
detailed balance. We aim to move the particle to the right
(against the force). For this purpose the controller measures the
particle location and consecutively raises from left to right the
reflecting barriers to trap the particle further and further to
the right. The next barrier to the right is raised when the
measurement indicates that the particle has crossed to the
righthand side. This implies that when the particle moves to the
left until the raised barrier location it finds a reflecting
boundary condition, while the particle has no bounds to its
displacements to the right.
\par

This defines a deterministic feedback control that pumps the
particle by using information about the location of the jumping
particle. We stress that a blind open-loop control strategy for
the lifting of the barriers cannot achieve direct flux against the
load. In addition, our closed-loop controller does not introduce
any extra energy in the system. Thus, the entropy reduction in the
system thanks to the information-gathering operation is the only
responsible for the pumping. In particular,
we highlight that a naive definition of efficiency as
$\eta=W/(-\Delta U_{\rm cont})$ is meaningless for engines that
work due to an information-dependent operation. Our general
results allow us to compute the maximum possible efficiency of
this pump as a case example, not only in the quasistatic regime
(large time intervals between two operations of the controller) but also when
it is operated non-quasistatically (for instance every time step).
\par

Let us first compute the maximum efficiency attainable when the
controller operates every time step. We consider the particle
initially at the origin with the reflecting barrier to the left
raised. At time $t_k$ the controller takes the value $C_k=1$ when the
next right barrier is raised or $C_k=0$ if the barrier remains
off. As the feedback control in this example satisfies the
deterministic condition~\eqref{error-free-cond}, the average
entropy reduction per step is given by Eq.~\eqref{SinfoRate}.
Furthermore, in order to simplify the computation of the entropy
rate, it is useful to change to a description in terms of a new  
stochastic process $\tilde C$, with $ \tilde C_s $ defined as the
number of steps between the raise of the barrier $s-1$ and that of
the barrier $s$ (first passage time). For example the event
$(C_1,\ldots,C_7)=(0,0,0,1,0,0,1)$ corresponds to the event
$(\tilde C_1,\tilde C_2)=(4,3)$. It is clear that we can establish
a one-to-one correspondence between $C$ and $\tilde C$, as both
represent univocally the control actions history. Calling $\langle
\tau \rangle$ as the average first passage time through the next
barrier position, we have that Eq.~\eqref{SinfoRate} reads
\begin{equation}
\lim_{t\to \infty} \frac{T\Sinfo}{t}
= \lim_{t\to \infty}\frac{-H(\past{t})}{t}=\lim_{s\to \infty}\frac{-H(
\tilde{\mathbf{C}}^{s})}{s \langle \tau \rangle}.
\end{equation}
[That is, $\bar H(\mathscr{C})=\bar
H(\tilde{\mathscr{C}})/\langle\tau\rangle$.] As the new tilde
variables are independent and identically distributed we have
$H(\tilde{\mathbf{C}}^{s})=sH(\tilde C_1)$. Thus,
\begin{equation}\label{escape}
    \lim_{t\to \infty} \frac{T\Sinfo}{t}
 =\frac{-H(\tilde C_1)}{\langle\tau\rangle}
 =\frac{\sum_{k=1}^{\infty}p_\tau (k)\ln p_\tau
    (k)}{\sum_{k=1}^{\infty}k p_\tau (k)},
\end{equation}
where $p_\tau (k)$ is the probability mass function of the first
passage time being $\tau=k$.
This asymptotic value, Eq.~\eqref{escape}, is reached in a
characteristic time $\langle \tau \rangle $. The probability $p_\tau (k)$
can be obtained from the transition probabilities between the
states of the jumping particle.
\par

On the other hand, the average potential increase is $W=f/\langle
\tau \rangle $. Therefore, the maximum efficiency attainable
at this nonquasistatic regime is obtained from Eq.~\eqref{eta} that reads
\begin{equation}\label{eq-eta}
  \eta_{\rm nq}= \frac{f}{H(\tilde C_1)}.
\end{equation}
\par

\subsection{One lattice site between consecutive barriers}

For instance, for the case with a single lattice site between two
barriers $p_\tau (k)= \alpha(1-\alpha)^{k-1}$, implying $H(\tilde
C_1)=H_b(\alpha)/\alpha$ and $\langle \tau \rangle=1/\alpha$.
Thus, the average entropy reduction per step is $H_b(\alpha)$, and
the average potential increase is $W=f/\langle
\tau \rangle=\alpha f$. Finally, the maximum efficiency attainable
at this nonquasistatic regime is $\eta_{\rm nq}=\alpha f/H_b(\alpha)$.
This result for the model
with a single site between two consecutive barriers can also be
obtained without using Eq.~\eqref{escape}. For this simple case
operation steps at different times are independent and $T\Delta
S_k=-H(C_k)$ with $p_{C_k}(1)=\alpha$. This gives an entropy
reduction per step $H_b(\alpha)$. On the other hand, the average
potential energy gain per step is $\alpha f$ because the particle
gains an energy $f$ with probability $\alpha$. In view of these
considerations we recover $\eta_{\rm nq}=\alpha f/H_b(\alpha)$.
\par

\subsection{Several lattice sites between consecutive barriers}

As $\alpha$ is the probability of jumping to the right, the
probability of the first passage time being $\tau=k$ is obtained
from the probability $p_{X_{k-1}}(n)$ of finding the particle at site $n$
(just to the left to the first barrier) at instant time 
$k-1$ as $ p_\tau (k)=\alpha  p_{X_{k-1}}(n) $. To evaluate this
probability we only need to know  the transition probabilities of
jumping between the different spatial positions (see
Fig.~\ref{fig:figure}). We shall call $\Pi$ as the matrix such that
its $(i,j)$th entry is the probability $p_{j\to i}$ of jumping
from the $j$ site to the $i$ site. Then, for the particle pump
with $n$ sites between barriers, $\Pi$ is the $n\times n$
tridiagonal matrix
\begin{equation}
  \Pi=
  \begin{pmatrix}
    1-\alpha & 1-\alpha &        &          & \\
    \alpha   &  0       & \ddots &          & \\
             & \alpha   & \ddots & 1-\alpha & \\
             &          & \ddots & 0        & 1-\alpha \\
             &          &        & \alpha   & 0
  \end{pmatrix}.
\end{equation}
Assuming that the particle is initially situated at the origin, the
probability $p_{X_{k-1}}(n)$ is given by the $(n,1)$th element of the
$(k-1)$th power of $\Pi$. Hence,
\begin{equation}
    p_\tau (k)=\alpha \Pi^{k-1}(n,1).
\end{equation}
For instance, for $n=1$ we recover $p_\tau (k)=\alpha
(1-\alpha)^{k-1}$, with $\alpha=1/(1+e^{f})$. For $n=2$ we get,
after some straightforward calculus, $ p_\tau (k) =
a(b_+^{k-1}-b_-^{k-1})$, where $ a := \alpha^2
/\sqrt{1+2\alpha-3\alpha^2}$ and $ b_{\pm} :=
(1-\alpha\pm\sqrt{1+2\alpha-3\alpha^2})/2$, with
$\alpha=1/(1+e^{f/2})$.
\par

\begin{figure}
    \includegraphics [scale=0.6] {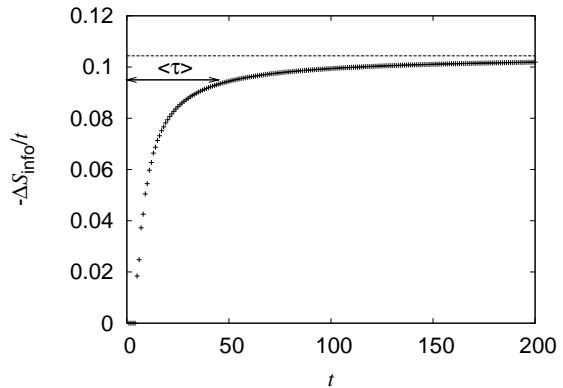}
    \caption{
Average entropy reduction per time step as a function of time for
the particle pump with $n=5$ lattice sites between barriers:
numerical simulations (+ signs) and asymptotic value (dashed
line). The asymptotic value is approached in a characteristic time
of the order of the mean first passage time $\langle \tau
\rangle$. Force $f=1$. Units $k_BT=1$ and $L=1$. }
    \label{fig:sinfo}
\end{figure}

Once the probabilities  $p_\tau (k)$ are obtained, the entropy reduction and
the efficiency can be computed with Eqs.~\eqref{escape} and~\eqref{eq-eta}
respectively. We plot in Fig.~\ref{fig:sinfo} this entropy reduction
$\lim_{t\to \infty} T\Sinfo / t $ for the particle pump with $n=5$
lattice sites between barriers, together with the time dependence of the
average 
entropy reduction per time step obtained by means of computer simulations
of the dynamics in the maximum measurement regime. As expected, this time
evolution tends to the theoretical asymptotic value in a characteristic time
of order $\langle \tau \rangle=\sum_{k=1}^{\infty} k p_\tau (k)$.

The numerical results in Fig.~\ref{fig:sinfo} have been obtained
evolving the particle distribution according to the known
transition probabilities. The entropy reduction in each
measurement is given by the entropy difference between the
particle distributions before and after the measurement. After the
measurement we keep one of the two possible particle distributions
chosen randomly with the probability of the corresponding
measurement output, and we evolve this particle distribution until
the next measurement. Following this procedure we have performed
several realizations of the control actions history, and
thereafter we have performed an average over realizations to
obtain the average entropy per time step as a function of time.
For these simulations we have considered $n=5$ lattice sites and
force $f=1$ (in units of $k_BT=1$ and $L=1$) or equivalently
$\alpha=1/(1+e^{1/5})\approx 0.45$.

\subsection{Quasistatic regime}

To conclude the analysis of the illustrating example, the
Markovian particle pump, we shall compute its maximum efficiency
in the quasistatic regime. Consider again the particle initially
situated at the origin. As the time between measurements is large
enough, the system has reached equilibrium when the controller
measures at a time $t\gg 1$. Hence
$p_{X_t}(m)=(1-e^{-f/n})e^{-fm/n}$ and the jumping particle is at
the righthand side of the next barrier with probability
$\sum_{m>n}p_{X_t}(m)=e^{-f}$. On the other hand, when the barrier
is raised the system gains a potential energy $f$. Thus, the
entropy reduction due to information is $H_b(e^{-f})$, while the
potential energy gained in one step is $fe^{-f}$. Therefore the
maximum efficiency for the quasistatic operation of the Markovian
particle pump is $\eta_{\rm q}=f e^{-f}/H_b(e^{-f})$. We note that
$0<\eta_{\rm nq}<\eta_{\rm q}<1$, as expected.
\par
In order to compare with results in Fig.~\ref{fig:sinfo} note that
for the same parameter values a measurement step in the
quasistatic regime reduces the entropy on average an amount
$H_b(e^{-1}) \approx 0.66 $. However, a measurement step in the
quasistatic regime requires many evolution time steps resulting in
a very low entropy reduction per time step.

\section{Conclusions} \label{sec:conclussions}

In this paper we have addressed the thermodynamics of closed-loop
controlled systems, focusing on what characterizes them, namely,
the use of information. Our results show explicitly how to
calculate the entropy reduction due to information,
Eq.~\eqref{Sinfo} or~\eqref{Sinfo2}. Therefore, they allow one to
compute the thermodynamic quantities and their relations for
feedback controlled systems. In particular, we have calculated the
thermodynamic relations for isothermal feedback controlled
systems, Eqs.~\eqref{isofirst}--\eqref{isocomb}, and also the
maximum efficiency attainable, Eqs.~\eqref{eta_def}
and~\eqref{eta}. As a case example, we have shown how to apply our
general results to a simple system that performs an isothermal
information-fueled particle pumping, for both a maximum
measurement regime and a quasistatic regime. The results presented in this
paper allow one to study the thermodynamics of many other feedback controlled
systems. It will be particularly interesting to obtain the thermodynamics of
feedback flashing ratchets that have been studied theoretically
\cite{ffrTheo}, and recently realized experimentally \cite{ffrExp}.
\par

\acknowledgments

We are grateful to Martin Bier for a critical reading of the paper.
We acknowledge financial support from MCYT (Spain) through the Research
Project No. FIS2006-05895, from the ESF Programme STOCHDYN, and from
UCM and CM (Spain) through Grant No. CCG07-UCM/ESP-2925.

\end{document}